\documentclass[oneside,11pt,a4paper]{article}
\usepackage{amssymb,amsmath,amsfonts,amsthm,epsfig}
\usepackage{rotating}
\theoremstyle{definition}
\newtheorem{problem}{Problem}
\newtheorem{myDefinition}{Definition}

\begin{document}
\normalsize
\setcounter{page}{1}%
\begin{center} \textbf{\uppercase{Algorithms of Fast Search\\[2mm] of Center, Radius and Diameter\\[2mm]on Weighted Graphs}}
\end{center}

\begin{center}{Urakov\,A.\,R., Timeryaev\,T.\,V.}
\end{center}

\begin{abstract}{Two problems in the search of metric characteristics on weighted undirected graphs with non-negative edge weights are being considered. The first problem: a weighted undirected graph with non-negative edge weight is given. The radius, diameter and at least one center and one pair of peripheral vertices of the graph are to be found. In the second problem we have additionally calculated the distances matrix. For the problems being considered, we proposed fast search algorithms which use only small fraction of graph's vertices for the search of the metric characteristics. The proposed algorithms have been compared to other popular methods of solving problems considered on various inputs.}\end{abstract}
\section*{Introduction}
\hspace*{\parindent} The metric (or numerical) characteristics of a graph are parameters defined by the shortest paths between vertices: center, radius and diameter. The diameter of a graph is the longest of the shortest paths between all pairs of vertices in the graph. The center is a vertex with the smallest maximum distance to any other vertex in the graph. This smallest distance is the radius of the graph.

These characteristics are essential properties of graphs, they represent valuable information about structures corresponding to the graphs. For example, in the facility location problem, the diameter can represent the maximum length of a waiting period, center and radius "---are the best location of the facility and the maximum time needed to reach this facility. In cases where graphs correspond to computer networks, the diameter can represent a connection speed between the two slowest nodes of the network and the radius can display the value of a delay between a server and the slowest node etc.

In general, the method of searching for these characteristics is to find the shortest paths between all pairs of vertices of a given graph and to determine distances and vertices which satisfy the definitions of the center, radius and diameter. Known algorithms performing this task on weighted graphs with $n$ vertices have a complexity from $\tilde{O}(Cn^{2,376})$ for graphs with integer edge weights less than $C$ \cite{ShoZwi99} and up to $O(n^{3}/\log^{2}n)$ for graphs with arbitrary edge weights \cite{cha07}. Obviously, the usage of these methods for structures consisting of hundreds of thousands to millions of vertices in many cases is unsuitable or even practically unfeasible due to a large amount of computing time.

Additionally we would like to mention the case where the shortest paths between all pairs of a graph's vertices have been calculated while the numerical characteristics are unknown and are to be found. This happens, for example, when initially the computing of numerical characteristics is not required or it is necessary to find characteristics not for the whole graph but for it's various subgraphs. Moreover, the structures of the subgraphs may change over time and, therefore, it is necessary to periodically recompute the values of the center, radius and diameter. A trivial solution in this case is the examination of all calculated shortest paths, it has a complexity $O(n^{2})$. The solution of this problem on graphs with large number of vertices or in cases where multiple recomputing of graph's characteristics is required may also have a long computing time.

In this paper, two metric characteristics problems are considered. Problem 1: is to find the center, radius and diameter of a graph defined only by the set of vertices $V$ and the set of edges $E$ and Problem 2: is to find the center, radius and diameter of a graph using additional information in the form of the calculated shortest paths between all pairs of the graph's vertices.

The problems are considered on connected weighted undirected graphs with non-negative real edge weights. Unconnected graphs are not considered because, by definition, the radius, center and diameter for them are infinite. In cases where characteristics for unconnected graphs are defined as some function of characteristics of its connected components (e.g., maximum), the problem can be reduced to the considered one by finding the connected components with the usage of fast existing methods \cite{knu97}. Only graphs with non-negative edge weights are considered since any graph without negative-weight cycles can be associated with the graph having only non-negative edge weights with preservation of the shortest paths \cite{joh77}.

In this paper, the algorithms for a quick search of metric characteristics of graphs for two problems mentioned above are presented. The main advantage of the proposed algorithms is that it is not necessary to examine all vertices of a graph to determine it's metric characteristics. The proposed algorithms have shown a significant reduction in computation time on weighted graphs of a different nature compared with popular methods in solving the problems considered.

\section{Problem definition}
\hspace*{\parindent} A connected undirected weighted graph $G=(V,E,w)$ is considered, here $V=(v_{1},v_{2},\dots,v_{n})$ is the set of vertices of the graph, $E=(e_{1},e_{2},\\ \dots,e_{n})$ is the set of edges of the graph, $e_{i}\subseteq V\times V$ and $w:E\to \mathbb{R}^{+,0}$ is a  nonnegative real weighted function on edges. It is further assumed that the powers of the vertex and edge sets are equal to $|V|=n$, $|E|=m$. Let us introduce some notations and definitions.

Denoted by $w(i,j)$ is the weight of the edge from vertex $v_i$ to vertex $v_j$. The length of the shortest path from vertex $v_i$ to vertex $v_j$ is called the distance from $v_i$ to $v_j$ and is denoted by $m_{ij}$. A matrix consisting of the shortest paths between all pairs of vertices of a graph is called the distance matrix and is denoted by $M=(m_{ij})$. A graph is called connected, if there is a path connecting any two of its vertices, that is the distance between any pair of vertices is finite $m_{ij}<\infty$. A weighted graph is called undirected, if the weights of the edges between any pair of vertices are equal in both directions $w(i,j)=w(j,i),\forall i,j$. Metric characteristics of a graph are defined as follows.

\begin{myDefinition}
The \emph{eccentricity} $\varepsilon(v_i)$ of vertex $v_i$ is the maximum distance from $v_i$ to any other vertex in the graph, $\varepsilon(v_i)=\max_{j=\overline{1,n}}m_{ij}$.
\end{myDefinition}

\begin{myDefinition}
The \emph{radius of a graph} $r$ is the minimum eccentricity of any vertex in the graph $r=\min_{i=\overline{1,n}}\varepsilon(v_i)$. Vertex $c$ achieving this minimum is called a \emph{central vertex (the center)}, $c:\varepsilon(c)=r$.
\end{myDefinition}

\begin{myDefinition}
The \emph{diameter of a graph} $d$ is the longest distance between any pair of vertices in the graph $d=\max_{i,j=\overline{1,n}}m_{ij}$. \emph{Peripheral vertices} are vertices of a graph such that distances between them are equal to the diameter.

\end{myDefinition}

In general, a graph can have several central vertices and several pairs of peripheral vertices. Particularly, in a complete graph with identical edge weights each vertex will simultaneously be central and peripheral. Our aim is to find at least one center and one pair of peripheral vertices, because more than often, the knowledge of the radius and diameter is more valuable than the  information about the vertices on which they are achieved. In this paper, two graph metric characteristics problems are to be considered.

\begin{problem}
Given a connected undirected weighted graph $G=(V,E,w)$ with a nonnegative real weight function on edges $w:E\to \mathbb{R}^{+,0}$. Find the radius $r$, diameter $d$ and at least one center $c$ and a pair of peripheral vertices of the given graph.
\end{problem}

\begin{problem}
A connected undirected weighted graph $G=(V,E,w)$ with a nonnegative real weight function on edges $w:E\to \mathbb{R}^{+,0}$ is given and the distance matrix $M$ is calculated. Find the radius $r$, diameter $d$ and at least one center $c$ and a pair of peripheral vertices of the given graph.
\end{problem}

Let us introduce two more notions. The single source shortest paths problem (SSSP) is a problem in which the shortest paths from a given vertex $v_i$ (the source) to all other vertices in the graph $v_j\in V$ are to be found. By SSSP($v_i$) the search of the shortest paths from a vertex $v_i$ to all other vertices of a given graph is denoted. The set of pivotal vertices is a set of $p$ vertices of a given graph $P=(p_1,p_2,\dots,p_p)$ such that the information about distances from vertices of this set is used for computing metric characteristics.

\section{Related works}
\hspace*{\parindent} In a case where the distance matrix of a graph is unknown, the most common method of searching for the graph's metric characteristics is to solve the all-pairs shortest path problem (APSP). Most of the algorithms solving APSP problem aren't universal and they show a good solution speed for graphs only with a certain set of properties. In particular, there are algorithms for sparse graphs \cite{dij59}, for graphs with bounded integer edge weights \cite{ShoZwi99} etc. Although the complexity of solving APSP gradually decreases with the invention of new algorithms, this approach for the search of metric characteristics of weighted graphs still remains unpractical for graphs with a large number of vertices .

There are algorithms that find metric characteristics for graphs having a special organization and describing peculiar structures. Among these, for example, is the algorithm for finding of the diameter for Small World Networks graphs \cite{takkos11} or the algorithm for finding of the center and diameter for Benzenoid Systems' graphs \cite{chedra02}. To increase the speed of a solution, these algorithms use peculiarities of the respective graphs, and therefore the range of their effective application is rigidly restricted.

One of the types of metric characteristics problems on graphs is the approximate calculation of the center and diameter of a graph. In these kind of problems, distances between graph vertices are searched with some degree of error allowing for reduction of computation time significantly. Among such, for example, is the Approximate neighborhood function algorithm \cite{bolrosvig11} or \cite{berkas07}.

Apart from the search of exact and approximate characteristics of a graph, the search for the so-called graph effective characteristics is widely spread. The effective radius of a vertex $v$ is the 90th-percentile of all distances from $v$. The effective diameter of a graph is the minimum distance within which 90\% of all vertices are within reach of each other. Certain algorithms for such problems are proposed, for example, in \cite{kuntsoapp10}.

A trivial method of solving Problem 2 is to examine the distances between all pairs of a graph vertices and determine those satisfying the definitions of the radius and diameter. The solving of this problem can be accelerated if computed distances are stored in special data structures (e.g., in the sorted list). Otherwise, the construction of such data structures will have a time complexity exceeding the time complexity of the trivial method. Other published methods of solving Problem 2 are unknown.

\section{Fast search algorithms}
\hspace*{\parindent} In short, the main idea of the proposed algorithms is as follows. First, a few pivotal vertices are chosen in a certain way and distances from them to all other vertices of a graph are computed. Using this information and the definitions of graph metric characteristics, the bounds of the radius, the diameter and their estimations on current iteration are found. After that, vertices having distances out of the found bounds are discarded. If there are vertices left, the number of pivotal vertices is increased and the bounds and estimations for the radius and diameter are recomputed. The process continues until all vertices are discarded. Current estimations of the radius and the diameter are the desired characteristics.

\subsection{The center and radius search}
\hspace*{\parindent} The algorithms for finding the center and the radius for Problems 1 and 2 are almost identical. The only difference is in the possibility to use computed distances from a considered vertex. In Problem 2, already computed distances, stored in a column of the distance matrix, are simply examined. In Problem 1, these distances are computed by the SSSP algorithm. At the heart of the search speed-up lies the determination of the lower $r_l$ and upper $r_u$ bounds of the graph radius with the usage of pivotal vertices $P$ and the examination of center pretenders $?_i$, until the equality $r_l=r_u$ doesn't hold true.

The graph radius $r$ has the following property. If the distances from all vertices of a graph to an arbitrary vertex $v_i$ are computed and the minimal of them $m_{ji}=\min_{k=\overline{1,n},k\neq i}m_{ki}$ has been found, the graph radius is inferior than the lower bound $r_l$ equal to that minimal value $r\geq r_l=m_{ji}$. Indeed, since $m_{ij}$ is the minimum distance to $v_i$ from all other vertices, the eccentricity of the center $c$ are not lower than this lower bound $r_l=m_{ji}\leq m_{ci}\leq\varepsilon(c)=r$. Moreover, $r_l$ is the lower bound of the radius in the case where it is defined as the smallest of the longest distances from all vertices of a graph to a subgraph of any $p$ vertices. In this case $r_l=\min_{i=\overline{1,n}}(\max_{p\in P}m_{ip})$ and $r_l=\max_{p\in P}m_{jp}\leq \max_{p\in P}m_{cp}\leq \varepsilon(c)=r$.

On the other hand, the graph radius $r$ is bounded from above by the eccentricity of any vertex $r\leq r_u=\varepsilon(v_j)$. Thus, the graph radius $r$ lies in the range $r_l\leq r\leq r_u$ and
\begin{equation}\label{eq3.1}
\text{if } r_l=r_u, \text{then } r=r_l=r_u
\end{equation}
Thereby, the graph radius $r$ and one of its centers can be found by iterative increasing the radius lower bound $r_l$ and decreasing the radius upper bound $r_u$ until the equality $r_l=r_u$ is held. Evidently, the speed of the convergence of $r_l$ and $r_u$ to the radius depends on the selection of center pretenders $c_i$ and pivotal vertices.

For the fast search of the radius, vertices "---center pretenders $c_i$ need to be found with as large as possible
\begin{equation}\label{eq3.2}
\max_{p\in P}m_{c_ip}=r_l=\min_{i=\overline{1,n}}(\max_{p\in P}m_{ip})
\end{equation}
and with as small as possible eccentricity $\varepsilon(c_i)$, here $P$ "--- the set of pivotal vertices. It is also desirable that the number of center pretenders $c_i$ examined and vertices in $P$ be as small as possible, since it minimizes the number of SSSP solved in Problem 1 and the number of the distance matrix elements examined in Problem 2.

Since the central vertex of many graphs is often situated near the geometrical center (if it could have been found), a good strategy is to choose center pretenders having a similar location. One way to find such a selection is to choose vertices $c_i$ having the minimum longest distance to peripheral vertices of a graph  $c_i:\max_{d\in D}m_{c_id}=\min_{i=\overline{1,n}}(\max_{d\in D}m_{id})$, where $D$ is the set of peripheral vertices of the graph. This approach requires you to find the diameter and peripheral vertices of a graph and therefore is very computationally complex. However, one can use the following simple method to get a good approximation of peripheral vertices.

The most remote from each other vertices $p_1$ and $p_2$ are sought.
\begin{equation}\label{eq3.3}
p_1,p_2:m_{p_1p_2}=\max_{i=\overline{1,n}}m_{ip_1}=\max_{i=\overline{1,n}}m_{ip_2}
\end{equation}

In order to do that, an arbitrary vertex $d_1$ and a vertex $d_2$ defined as the most remote from $d_1$ $d_2:m_{d_1d_2}=\max_{i=\overline{1,n}}m_{d_1i}$ are chosen. Subsequent vertices in this <<sequence>> are defined similarly
\begin{equation}\label{eq3.4}
d_{j+1}:m_{d_jd_{j+1}}=\max_{i=\overline{1,n}}m_{d_ji}
\end{equation}
The process continues until the condition $d_{j-1}=d_{j+1}$ is held, in this case the desirable vertices have been found and $p_1=d_j$, $p_2=d_{j+1}$.

Vertices $p_1$ and $p_2$ are included in the set of pivotal vertices $P=\{p_1, p_2\}$. Center pretenders $c_i$ are determined by \eqref{eq3.2}. If for the first center pretender $c_1$ the condition $r_l=r_u$ is held, $c_1$ is the center of a graph and $r=r_l=r_u$. If $r_u>r_l$, the most remote from $c_1$ vertex $p_3:m_{c_1p_3}=\varepsilon(c_1)$ is added to the set of pivotal vertices $P$ and the search of $c_2$ is performed by \eqref{eq3.2}. The process continues until the condition \eqref{eq3.1} is held for some pretender.

By this approach, the search of the graph center and the radius has several advantages. Firstly, the computation (or examination for Problem 2) of the shortest paths takes place during the search, and if the condition \eqref{eq3.1} is held early enough, there is no need to perform the rest of the procedure for  most of the graph vertices. Secondly, it is not necessary to compute maximums in \eqref{eq3.2} for every new center pretender, one can store computed maximums and after adding a new vertex $p_i$ to $P$, compute only maximums from the stored values and the distances from a new vertex $p_i$. The description of the algorithm for the both problems is represented in fig.1
\smallskip

{\centering
\fbox{
\parbox{12cm}{
{\centering \textbf{The center and radius search algorithm (R1, R2)}}

{\raggedright Input: a graph $G=(V,E,w)$}

\hspace*{\parindent} distance matrix $M$  (\emph{only for Problem 2}).

\hspace*{\parindent} $r_l=0$, $r_u=\infty$

1. The search speed-up

\hspace*{\parindent} Solve SSSP($d_i$) for the vertices $d_i$ from \eqref{eq3.4} (\emph{only for Problem 1}).

\hspace*{\parindent} Determine $p_1,p_2$ by \eqref{eq3.3} and include them into $P$.

2. Find a center pretender

\hspace*{\parindent} \hspace*{\parindent} Determine the vertex $c_i$ and $r_l$ by \eqref{eq3.2}.

3. Verification of the pretender

\hspace*{\parindent} If the shortest paths aren't computed for $c_i$, solve SSSP($c_i$) (\emph{only for Problem 1}).

\hspace*{\parindent} Find the eccentricity $\varepsilon(c_i)$ and, if $\varepsilon(c_i)<r_u$, $r_u=\varepsilon(c_i)$.

\hspace*{\parindent} If $r_u\neq r_l$, add the vertex $p_j:m_{c_ip_j}=\varepsilon(c_i)$ to $P$ and go to the step 2.

\hspace*{\parindent} Otherwise, $c_i$ is one of the graph centers, $r=r_l$ is the graph radius,  the algorithm ends.}}}

{\raggedright Fig 1: The center and radius search algorithm. Additional data and actions for Problems 1 and 2 are indicated in italics.}

The time complexity of the algorithm, evidently, depends on the number of center pretenders $c_i$ examined. For graphs with different structures this number can vary widely, so it is impossible to rigorously evaluate the complexity for the case in general, nevertheless it is possible to roughly indicate the lower and upper bounds. Consider the best and the worst scenarios for Problem 1. In the best case, when the first center pretender $c_1$ is one of the graph centers, one has to solve the SSSP problem for $k$ vertices $d_i$ from the first step of the algorithm, determine $c_1$ by \eqref{eq3.2} and find the eccentricity of $c_1$ by solving SSSP($c_1$). The time complexity in this case is determined by $k\cdot diff(SSSP)+k\cdot O(n)$, where $diff(SSSP)$ is the time complexity in solving SSSP. In the theoretically worst case, when the radius is determined through the examination of all possible pretenders $c_i$, we obtain $n\cdot diff(SSSP)+O(n^2)$, here $O(n^2)$ is the complexity of the eccentricities search of all $n$ vertices and determination $c_i$ by \eqref{eq3.2}.

\subsection{The diameter search}
\hspace*{\parindent} The principal difference in the search for the diameter is that, unlike for the radius, it is impossible in general to determine a convenient upper bound for the diameter. This implies another difference from the radius search "--- the existence of two different algorithms for Problems 1 and 2. Without an upper bound for the diameter the number of pairs of vertices examined $v_i, v_j$ "---pretenders on graph peripheral vertices "--- can be large enough. Hence the examination of each vertex $v_k$ in Problem 1 requires a preliminary solution of SSSP($v_k$). Therefore, for Problem 1 it is reasonable to perform an auxiliary procedure allowing to reduce the number of pairs of vertices $v_i, v_j$ examined and having a smaller computational complexity compared to the solving of SSSP. The sorting of vertices pairs $v_i, v_j$ by a potential distance between them and the examination of the sorted pairs in descending order are chosen in the capacity of such procedure. In Problem 2, the sorting is impractical due to the high complexity of the sorting of an array in comparison with the examination of a few extra columns of the distance matrix.

Obviously, the graph diameter $d$ is bounded from below by the distance between any two of its vertices $d\geq d_l=m_{ij}, \forall i,j=\overline{1,n}$. The upper bound of a graph can not be determined without the examination of distances greater than the lower bound $m_{kl}>d_l$. That is, the upper bound can be computed only after the determination of the graph diameter and thus has no practical use. In such circumstances, one way to speed-up the search of the diameter is to find a large lower bound $d_l$ with a minimal computational cost and examine only pairs of vertices $v_k, v_l$ having distances between them potentially exceeding the lower bound $m_{kl}>d_l$.

The maximal distance between the pivotal vertices is used in the algorithms as the first approximation of the diameter's lower bound $d_l$. As mentioned before, the pivotal vertices are  quite a good approximation to graph peripheral vertices. The distances from these vertices to the graph center $c$ are used in the algorithms for the estimation of potential distances. The choosing of $c$ in that role allows us to discard from the examination a large number of vertices $v_k, v_l$ which are not peripheral $m_{kl}<d_l$.

The diameter search algorithms consist of two parts. At first, one of the graph centers is scanned by the algorithms described in the previous section. Then the direct search of the graph's diameter, based on the information about distances from the center to other vertices of the graph is performed. That is, the diameter search algorithms completely solve both considered problems and thus may be called the radius and diameter search algorithms.

\subsubsection{Algorithm for Problem 2}
\hspace*{\parindent} After one of the graph centers $c$ has been found, only the distances from vertices $v_i$ for which the following condition is held are examined
\begin{equation}\label{eq3.5}
m_{ic}>d_l/2
\end{equation}
where $d_l$ is the current lower bound of the diameter determined by
\begin{equation}\label{eq3.6}
d_l=\max_{i,j\in P}m_{ij}
\end{equation}
and $P$ is the set of pivotal vertices generated during the computation of the radius. Indeed, if the graph diameter is greater than the current lower bound and $v_i$ is one of graph peripheral vertices, for some vertex $v_j$ by the triangle inequality for the distance matrix, we have $m_{ic}+m_{cj}\geq m_{ij}=d>d_l$. That is, $m_{ic}+m_{cj}>d_l=2\cdot d_l/2$ and it is necessary either $m_{ic}>d_l/2$ or $m_{cj}>d_l/2$. This proves the necessity to consider only vertices $v_i$ satisfying \eqref{eq3.5}, by virtue of a symmetry of the matrix $M$. The description of the algorithm is shown on figure 2.

\smallskip

{\centering
\fbox{
\parbox{12cm}{
{\centering \textbf{The diameter search algorithm for Problem 2 (D2)}}

{\raggedright The input: a graph $G=(V,E,w)$}, the distance matrix $M$.

\hspace*{\parindent} $d_l=0$

1. The search of the graph center

\hspace*{\parindent} Find one of the graph centers $c$ by using the algorithm from section 3.1.

\hspace*{\parindent} Compute the lower bound of the diameter $d_l$ by \eqref{eq3.6}.

2. The search of the diameter

\hspace*{\parindent} Examine the distances from vertices $v_i$ satisfying \eqref{eq3.5} may be the diameter. If $m_{ij}>d_l$, $d_l=m_{ij}$.

The output: $d_l$ is the graph diameter.
}
}
}

{\raggedright Fig 2: The diameter search algorithm for Problem 2.}

\subsubsection{Algorithm for Problem 1}
\hspace*{\parindent} In Problem 1 the examination of all vertices $v_i$ satisfying \eqref{eq3.5} can be quite long due to the necessity to solve SSSP($v_i$) for each such vertex. Therefore, to narrow the range of peripheral vertex pretenders one can use the information involving distances from two vertices to the graph center.

By definition of the shortest path, elements of the distance matrix $M$ satisfy the triangle inequality
\begin{equation}\label{eq3.7}
m_{ij}\leq m_{ik}+m_{kj},\forall i,j,k
\end{equation}
Therefore, after having computed distances from an arbitrary vertex $v_v$ to all other vertices, the distances only between pairs of vertices $v_k, v_l$ satisfying the following condition have to be examined
\begin{equation}\label{eq3.8}
m_{kv}+m_{vl}>d_l
\end{equation}
For the vertices not satisfying this inequality, $m_{kl}\leq m_{kv}+m_{vl}\leq d_l$, i.e. the distances between them are not greater than the current lower bound. As the vertex $v_v$ the graph center $c$ is used in the algorithm.

According to \eqref{eq3.7}, vertices $v_k, v_l$ with the maximal sum $m_{kv}+m_{vl}$ having the maximal potential distance. Therefore, it is reasonable to examine vertices $v_k, v_l$ in descending order of the value $m_{kv}+m_{vl}$. If during this "top-down" passage for some pair of vertices $v_k, v_l$ inequality \eqref{eq3.8} isn't held, the examination of the remaining vertices with smaller potential distances between them is not needed.

The sorting of $n^2$ pairs $v_k, v_l$ can be quite long compared to the total time of the diameter search, thus one can use the following method which allows reduction of the computation time while somewhat degrading the quality of the sort. Instead of the sorting of pairs $v_k, v_l$ by the sum $m_{kc}+m_{cl}$ one should sort vertices $v_i$ by the distance to the center $m_{ic}$ and perform verification of condition \eqref{eq3.8} for vertices pairs in the following order

\begin{equation}\label{eq3.9}
v_{m_1}v_{m_2},v_{m_1}v_{m_3},\dots,v_{m_1}v_{m_n},v_{m_2}v_{m_3},v_{m_2}v_{m_4}\dots
\end{equation}
Here $m_i$ is the number of the vertex with the $i$th largest distance to the graph center. If some $v_{m_i}v_{m_j}$ are not satisfying \eqref{eq3.8}, the pair $v_{m_{i+1}}v_{m_{i+2}}$ is examined. If \eqref{eq3.8} isn't held and $j=i+1$, the algorithm is terminated and $d_l$ is the graph diameter. The description of the algorithm is shown in figure 3.

\smallskip

{\centering
\fbox{
\parbox{12cm}{
{\centering \textbf{The diameter search algorithm for Problem 1 (D1)}}

{\raggedright The input: a graph $G=(V,E,w)$}.

\hspace*{\parindent} $d_l=0$

1. The search of the graph center

\hspace*{\parindent} Find one of the graph centers $c$ using the algorithm from section 3.1.

\hspace*{\parindent} Compute the lower bound of the diameter $d_l$ by \eqref{eq3.6}.

2. Sorting

\hspace*{\parindent} Sort vertices $v_i$ by the distance to the center $c$ in descending order.

3. The examination of pretenders

\hspace*{\parindent} Choose a pair of pretenders from list \eqref{eq3.9}

\hspace*{\parindent} If pair $v_k, v_l$ satisfies \eqref{eq3.8}, then if the shortest paths for $v_k, v_l$ are not computed, solve SSSP($v_k$) and SSSP($v_l$).

\hspace*{\parindent} If $m_{kl}>d_l$, $d_l=m_{kl}$

The output: $d_l$ is the graph diameter.}}
}

{\raggedright Fig 3: The diameter search algorithm for Problem 1.}

As in the case with the search of the radius, the complexity of the algorithm heavily depends on the graph structure and on how fortunately the pivotal vertices were chosen. In general, the complexity of the algorithm can be estimated by $diff(Ar)+diff(Sort(n))+k\cdot diff(SSSP)+O(x)$. Where $diff(Ar)$ is the complexity of the radius and the center search algorithm, $diff(Sort(n))$ is the complexity of the used sorting algorithm, $k\cdot diff(SSSP)$ is the complexity of the solving of SSSP for $k$ vertices in step 3 of the algorithm and $O(x)$ is the examination of $x$ pair of vertices $v_k, v_l$ to satisfy inequality \eqref{eq3.8}.

\section{Tests results}
\hspace*{\parindent} All the tests have been performed on a computer equipped with Intel Core i5 530 (2,93 GHz) CPU and 3 GBs of RAM on the 32-bit edition of Windows XP. The source code has been written on C++ programming language in Borland C++ Builder 6 IDE. The first set of the data used is weighted graphs of the USA road networks from the public source \cite{dimURL}. The second set of the data used is complete graphs with random edge weights generated by us. For complete graphs results are taken as an average over 10 runs on different graphs with the same dimension. The parameters of the tested graphs are presented in table 1.

\begin{table}[h]
\caption{The parameters of the tested graphs.}
\label{tab4.1}
\begin{center}
\begin{tabular}{|c|c|c|c|} \hline
Name & Vertices & Edges & Average degree \\ \hline
RN\_1 & 1001 & 2432 & 2,43 \\
\hline
RN\_2 & 2007 & 5288 & 2,63 \\
\hline
RN\_5 & 5000 & $1,34\cdot10^4$ & 2,69 \\
\hline
RN\_7 & 7000 & $1,8\cdot10^4$ & 2,58 \\
\hline
RN\_10 & $10^4$ & $2,7\cdot10^4$ & 2,69 \\
\hline
RN\_15 & $1,5\cdot10^4$ & $4,38\cdot10^4$ & 2,92 \\
\hline
RN\_20 & $2\cdot10^4$ & $5,41\cdot10^4$ & 2,71 \\
\hline
RN\_NY & $2,64\cdot10^5$ & $7,34\cdot10^5$ & 2,78 \\
\hline
RN\_BAY & $3,21\cdot10^5$ & $8\cdot10^5$ & 2,49 \\
\hline
RN\_COL & $4,36\cdot10^5$ & $1,06\cdot10^6$ & 2,43 \\
\hline
RN\_FLA & $1,07\cdot10^6$ & $2,71\cdot10^6$ & 2,53 \\
\hline
RN\_NW & $1,21\cdot10^6$ & $2,84\cdot10^6$ & 2,35 \\
\hline
RN\_LKS & $2,76\cdot10^6$ & $6,89\cdot10^6$ & 2,5 \\
\hline
RN\_E & $3,6\cdot10^6$ & $8,78\cdot10^6$ & 2,44 \\
\hline
RN\_W & $6,26\cdot10^6$ & $1,52\cdot10^7$ & 2,43 \\
\hline
F\_1 & 1000 & $10^6$ & 1000 \\
\hline
F\_2 & 2000 & $4\cdot10^6$ & 2000 \\
\hline
F\_5 & 5000 & $2,5\cdot10^7$ & 5000 \\
\hline
F\_7 & 7000 & $4,9\cdot10^7$ & 7000 \\
\hline
F\_10 & $10^4$ & $10^8$ & $10^4$ \\
\hline
\end{tabular}
\end{center}
\end{table}

The proposed algorithms for Problem 1 (denoted by R1 and D1) have been compared to the a) Dijkstra algorithm performed for each vertex of graph in the implementation with a binary heap \cite{cor01} for graphs with a small average vertex degree and b) to the Floyd-Warshall algorithm \cite{flo62} for complete graphs (denoted by RC1 and DC1). As it is impossible within reasonable time to obtain the solution of the problem on high-dimensional graphs by using the Dijkstra algorithm, the running time for graphs from RN\_BAY onward has been approximated using the least squares method. The data used for the approximation are subgraphs of different dimensions of the tested graphs, a part of them is represented in table \ref{tab4.1} (RN\_1, RN\_2 etc.). The running time of the Dijkstra algorithm for all vertices of a graph has been approximated by $7,1\cdot10^{-7}n^2\log(n)-1,1\cdot10^{-7}nm\log(n)+0,66$ for the search of the radius and by $7,2\cdot10^{-7}n^2\log(n)-1,1\cdot10^{-7}nm\log(n)+0,98$ for the search of the diameter. The Dijkstra algorithm in the implementation with a binary heap has been used for the solving of SSSP in the proposed algorithms. Quicksort has been used as the sorting algorithm in D1. The results of the tests for Problem 1 are shown in table \ref{tab4.2}.

\begin{sidewaystable}[h]
\caption{The tests results for problem 1. R1 SSSP, pieces is the number of the SSSP solving during R1, R1 SSSP, share is the share of vertices with solved SSSP of the total number of the graph vertices. Similarly for D1. Approximate results denoted by $^*$.}
\label{tab4.2}
\begin{center}
\begin{tabular}{|c|c|c|p{0.8cm}|p{1.8cm}|c|c|p{0.8cm}|p{1.8cm}|} \hline
Graph & RC1, sec & R1, sec & R1 SSSP, pieces & R1 SSSP, share & DC1, sec & D1, sec & D1 SSSP, pieces & D1 SSSP\newline, share \\ \hline
RN\_1 & 1,33 & 0,016 & 4 & 0,004 & 1,31 & 0 & 4 & 0,004 \\
\hline
RN\_2 & 5,45 & 0,031 & 7 & 0,0035 & 5,45 & 0,031 & 10 & 0,005 \\
\hline
RN\_5 & 39 & 0,031 & 4 & 0,0008 & 39 & 0,61 & 72 & 0,0144 \\
\hline
RN\_7 & 75 & 0,063 & 6 & 0,00086 & 75 & 5,5 & 492 & 0,0702 \\
\hline
RN\_10 & 170 & 0,17 & 10 & 0,001 & 170 & 0,72 & 41 & 0,00,1  \\
\hline
RN\_15 & 373 & 0,17 & 7 & 0,00047 & 374 & 0,77 & 30 & 0,002 \\
\hline
RN\_20 & 761 & 0,23 & 6 & 0,0003 & 762 & 25 & 635 & 0,0318 \\
\hline
RN\_NY & $1,54\cdot 10^5$  & 3,5 & 6 &$ 2,3\cdot 10^{-5}$  & $1,54\cdot 10^5$  & 5,3 & 9 & $3,4\cdot 10^{-5}$  \\
\hline
RN\_BAY & $2,49\cdot 10^{5*}$ & 3,16 & 4 & $1,3\cdot 10^{-5}$  & $2,49\cdot 10^{5*}$ & 3,31 & 4 & $1,25\cdot 10^{-5}$  \\
\hline
RN\_COL & $4,76\cdot 10^{5*}$ & 7,09 & 6 & $1,4\cdot 10^{-5}$  & $4,77\cdot 10^{5*}$ & 233 & 198 & 0,00045 \\
\hline
RN\_FLA & $2,99\cdot 10^{6*}$ & 13 & 4 & $3,7\cdot 10^{-6}$  & $2,99\cdot 10^{6*}$ & 16 & 5 & $4,7\cdot 10^{-6}$  \\
\hline
RN\_NW & $4,03\cdot 10^{6*}$ & 14 & 4 & $3,3\cdot 10^{-6}$  & $4,03\cdot 10^{6*}$ & 15 & 4 & $3,3\cdot 10^{-6}$  \\
\hline
RN\_LKS & $2,14\cdot 10^{7*}$ & 37 & 4 & $1,5\cdot 10^{-6}$  & $2,14\cdot 10^{7*}$ & 38 & 4 & $1,5\cdot 10^{-6}$  \\
\hline
RN\_E & $3,76\cdot 10^{7*}$ & 49 & 4 & $1,1\cdot 10^{-6}$  & $3,77\cdot 10^{7*}$ & 51 & 4 & $1,1\cdot 10^{-6}$  \\
\hline
RN\_W & $1,18\cdot 10^{8*}$ & 92 & 4 & $6,4\cdot 10^{-7}$  & $1,19\cdot 10^{8*}$ & 97 & 4 & $6,4\cdot 10^{-7}$  \\
\hline
F\_1  & 10 & 0,18 & 9 & 0,009 & 10 & 0,27 & 13,9 & 0,0139 \\
\hline
F\_2  & 81 & 1 & 10,3 & 0,0052 & 81 & 1,75 & 17,8 & 0,0089 \\
\hline
F\_5  & 1246 & 12 & 9,5 & 0,0019 & 1246 & 21 & 15,4 & 0,0031 \\
\hline
F\_7  & 3401 & 33 & 10 & 0,0014 & 3401 & 45 & 13,7 & 0,002 \\
\hline
F\_10  & 9853 & 71 & 9,5 & 0,00095 & 9853 & 123 & 16,5 & 0,0017 \\
\hline
\end{tabular}
\end{center}
\end{sidewaystable}

On all the tested graphs a significant reduction in search time of the radius and diameter for Problem 1 is observed when the proposed algorithms are used. The percentage of vertices with solved SSSP doesn't exceed 0,9\% for the radius search and 7\% for the diameter search.

For Problem 2 the tests have been performed on graphs with the number of vertices not exceeding $10^4$ due to the infeasibility of storing the distances information in a computer memory for larger graphs. The proposed algorithms (R2, D2) have been compared to the simple examination of all elements of the distance matrices of graphs for the search of the radius (RC2) and to the examination of the upper triangle of distance matrices for the search of the diameter (DC2). Due to a small solution time for the problem on the tested graphs, each algorithm has been run 1,000 times consecutively. The tests results for Problem 2 are shown in table \ref{tab4.3}.

\begin{sidewaystable}[h]
\caption{The tests results for Problem 2.}
\label{tab4.3}
\begin{center}
\begin{tabular}{|c|c|c|c|c|c|c|} \hline
Graph & RC2, sec & R2, sec & R, speed-up, times & DC2, sec & D2, sec & D, speed-up \\ \hline
RN\_1  & 4,21 & 0,031 & 135 & 1,7 & 0,047 & 36 \\
\hline
RN\_2  & 156 & 0,14 & 1114 & 7,3 & 0,17 & 42 \\
\hline
RN\_5  & 95 & 0,16 & 593 & 45 & 1,38 & 32 \\
\hline
RN\_7  & 181 & 0,38 & 476 & 87 & 16 & 5 \\
\hline
RN\_10  & 374 & 0,89 & 420 & 177 & 2,19 & 80 \\
\hline
F\_1  & 4,07 & 0,072 & 56 & 2,08 & 0,107 & 19 \\
\hline
F\_2  & 16 & 0,16 & 100 & 8,16 & 0,26 & 31 \\
\hline
F\_5  & 100 & 0,43 & 232 & 50 & 0,69 & 72 \\
\hline
F\_7  & 196 & 0,63 & 311 & 98 & 0,86 & 113 \\
\hline
F\_10  & 399 & 0,81 & 492 & 200 & 1,27 & 157 \\

\hline
\end{tabular}
\end{center}
\end{sidewaystable}

For problem 2 a significant advantage of the proposed algorithms is also observed "--- the speed-up of the radius search from 56 times onward, the speed-up of the diameter search from 5 times onward.

\section*{Conclusion}
\hspace*{\parindent} The proposed algorithms solve the stated metric characteristics Problems 1 and 2 using the distance information only of a small fraction of graph vertices. The tests results have shown that, at least on the tested graphs, the percentage of the used information decreases with increas of dimensions of graphs. This behavior of the algorithms indicates that high-dimensional graphs have greater gaining time.

The main results of the performed tests in terms of numbers are follows. For the search of the center and radius for Problem 1, the number of vertices with solved SSSP doesn't exceed 0,9\%, the running time speed-up changes in the range from 55 up to $1,28\cdot 10^6$ times faster. For the diameter search for Problem 1 these figures are, accordingly, 7\% and from 14 up to $1,22\cdot 10^6$ times faster. For Problem 2 the range of speed-up changes from 56 up to $1,14\cdot 10^3$ for the search of the center and radius and from 5 up to 157 times faster for the diameter search.

In further research on the topic of this paper, the test of the proposed algorithms on other sorts of graphs and estimation of the sphere of application of the algorithms can be fulfilled. Another direction of research is the further optimization of the algorithms through the usage of faster sorting algorithms and the usage of algorithms of solving the SSSP relevant for considered graphs. The question of applicability of the approach used for the algorithms in search of all graph peripheral vertices and centers also seems interesting.


\begin{thebibliography}{25}
\bibitem{berkas07}
Berman\,P., Kasiviswanathan\,S.\,P.
{\rm Faster Approximation of Distances in Graphs // Algorithms and Data Structures. Lecture Notes in Computer Science --- 2007 --- Vol. 4619. --- P.~541--552}
\bibitem{bolrosvig11}
Boldi\,P., Rosa\,M., Vigna\,S.
{\rm HyperANF: approximating the neighbourhood function of very large graphs on a budget // Proceedings of the 20th international conference on World wide web. --- New York: ACM, 2011. --- P.~625--634}
\bibitem{cha07}
Chan\,T.\,M.
{\rm More Algorithms for All-Pairs Shortest Paths in Weighted Graphs // Proceedings of the thirty-ninth annual ACM symposium on Theory of computing. --- New York: ACM, 2007. --- P.~590--598}
\bibitem{chedra02}
Chepoi\,V., Dragan\,F. et al.
{\rm Center and Diameter Problems in Plane Triangulations and Quadrangulations // Proceedings of the thirteenth annual ACM-SIAM symposium on Discrete algorithms. --- Philadelphia: Society for Industrial and Applied Mathematics, 2002. --- P.~346--355}
\bibitem{cor01}
Cormen\,T.\,H., Leiserson,\,C.\,E., Rivest\,R.\,L., Stein\,C.
{\rm Introduction to Algorithms, Second Edition ---MIT Press, 2001. --- 1202~p.}
\bibitem{dij59}
Dijkstra\,E.\,W.
{\rm A note on two problems in connection with graphs //  Numerische Mathematik 1. --- 1959.  --- P.~83--89}
\bibitem{flo62}
Floyd\,R.\,W.
{\rm Algorithm 97: Shortest Path // Communications of the ACM --- 1962. --- Vol. 5, N 6. --- P.~345}
\bibitem{joh77}
Johnson\,D.\,B.
{\rm Efficient algorithms for shortest paths in sparse networks // Journal of the ACM. --- 1977. --- Vol. 24, N 1. --- P.~1--13}
\bibitem{knu97}
Knuth\,D.\,E.
{\rm The Art Of Computer Programming Vol 1. 3rd ed., --- Boston: Addison-Wesley, 1997. --- 650~p.}
\bibitem{kuntsoapp10}
Kung\,U., Tsourakakis\,C.\,E., Appel\,A.\,P., Faloutsos\,C., Leskovec\,J.,
{\rm Radius Plots for Mining Tera-byte Scale Graphs: Algorithms, Patterns, and Observations. // SIAM International Conference on Data Mining 2010 --- 2010 --- P.~548--558}
\bibitem{ShoZwi99}
Shoshan\,A., Zwick\,U.
{\rm All Pairs Shortest Paths in Undirected Graphs with Integer Weights // Proceedings of the 40th Annual Symposium on Foundations of Computer Science. --- Washington: IEEE Computer Society, 1999. --- P.~605--614}
\bibitem{takkos11}
Takes\,F.\,W., Kosters \,W.\,A.
{\rm Determining the diameter of small world networks // Proceedings of the 20th ACM international conference on Information and knowledge management. --- New York: ACM, 2011.  --- P.~1191--1196}
\bibitem{dimURL}
{\rm 9th DIMACS Implementation Challenge "---Shortest Paths. URL: http://www.dis.uniroma1.it/challenge9/download.shtml (access date: 16.07.2012).}


\end{thebibliography}
\end{document}